\documentclass[authoryear,preprint,review,12pt]{elsarticle}

\usepackage{amssymb,amsmath}
\usepackage{latexsym}
\usepackage{amsmath}
\usepackage{graphicx}

\journal{JMPS}

\begin{document}

\begin{frontmatter}



\title{Bounds on Effective Dynamic Properties of Elastic Composites}


\author{Sia Nemat-Nasser and Ankit Srivastava}

\address{Department of Mechanical and Aerospace Engineering\\ 
University of California, San Diego\\
La Jolla, CA, 92093-0416 USA}

\begin{abstract}

We present general, computable, improvable, and rigorous bounds for the total energy of a finite heterogeneous volume element $\Omega$  or a periodically distributed unit cell of an elastic composite of any known distribution of inhomogeneities of any geometry and elasticity, undergoing a harmonic motion at a fixed frequency or supporting a single-frequency Bloch-form elastic wave of a given wave-vector. These bounds are rigorously valid for \emph{any consistent boundary conditions} that produce in the finite sample or in the unit cell, either a common average strain or a common average momentum.  No other restrictions are imposed.  We do not assume statistical homogeneity or isotropy. Our approach is based on the Hashin-Shtrikman (1962) bounds in elastostatics, which have been shown to provide strict bounds for the overall elastic moduli commonly defined (or actually measured) using uniform boundary tractions and/or linear boundary displacements; i.e., boundary data corresponding to the overall uniform stress and/or uniform strain conditions. Here we present strict bounds for the dynamic frequency-dependent constitutive parameters of the composite and  give explicit expressions for a direct calculation of these bounds.  

\end{abstract}

\begin{keyword}
Bounds \sep Effective Dynamic Properties \sep Metamaterials \sep Homogenization

\end{keyword}

\end{frontmatter}


\section{Introduction}

In micromechanics, one seeks  to estimate the overall (or effective) mechanical properties of a material
in terms of its microstructure and the properties of its microconstituents.
The overall properties are defined by
relating the unweighted volume average of the kinematical and dynamical field quantities, taken over a suitably large sample called the representative volume element ($\Omega$).
A well-established classical approach is to replace the heterogeneous composite by a homogeneous one and then introduce eigenstress or eigenstrain (polarization stress or strain) fields such that the stress and strain fields in the equivalent homogeneous solid coincide with the actual stress and strain fields of the original heterogeneous $\Omega$; see \cite{hashin1959moduli,hashin1962some,hashin1962variational,kröner1977bounds}.   This scheme leads to a set of integral equations, referred to as \emph{the consistency conditions}, which need to be solved to obtain the required exact homogenizing eigenstress and/or eigenstrain field.  While the homogenizing eigenfields will depend on the choice of the reference properties, the final results are unique and independent of that choice. Moreover, since this homogenization scheme is based on the volume average of the eigenfields, these averages can be calculated rather accurately using various approximations; see \cite{nemat1999micromechanics} for details and references. 

Since the microstructure of most materials is rather complex, bounds have been developed to estimate the overall properties of heterogeneous materials.  Among these the Hashin-Shtrikman (1962) variational principle and the resulting bounds for the overall parameters have been most extensively used to estimate the overall elastostatic  properties of
heterogeneous materials; \cite{willis1981avariational,willis1981bvariational}.  

The overall properties of a finite heterogeneous composite, as well as the corresponding bounds, in general will depend on the geometry and the prescribed boundary conditions. There are however two exact energy bounds that have allowed creating bounds for effective properties, which 
would be valid for any boundary data. For elastostatic problems, it has been shown (\cite{nemat1995universal}) that the elastic energy (complementary elastic energy) of a finite composite subjected to any boundary conditions is bounded by its corresponding energy under uniform tractions (linear displacements) provided that all considered boundary data produce the same volume-averaged strain (stress) in the composite. It was also shown that, in general, there are two \emph{universal bounds} for the two components of the overall modulus tensor, which depend only on the volume fraction but not on the detail distributions of the micro-constituents of the composite.  Note that since the measurement of the overall properties of materials are generally performed under essentially uniform boundary data, the Hashin-Shtrikman bounds for uniform boundary data provide powerful tools to guide the analysis of the experimental results.

Recent interest in the character of the overall dynamic properties of composites with tailored microstructure necessitates a systematic homogenization procedure to express the dynamic response of an elastic composite in terms of its average effective compliance and density. \cite{willis2009exact} has presented a homogenization method based on an ensemble averaging technique of the 'Bloch' reduced form of the wave propagating in a periodic composite; see also \cite{willis2011effective}.  A complementary micromechanical method to calculate the effective dynamic properties of general three-dimensional periodic elastic composites has been proposed by \cite{srivastava2011overall}, which generalizes the results for layered composites (\cite{nematnasser2011overall,nemat2011homogenization}). 
Furthermore, \cite{srivastava2011universal} presented universal theorems which are the dynamic analogues of the static theorems presented in \cite{nemat1995universal,nemat1999micromechanics} and have proven that, at a fixed frequency, the total elastodynamic energy (strain energy plus the kinetic energy), 
and the total complementary elastodynamic energy (complementary strain energy plus the kinetic energy) of 
$\Omega$, \emph{subjected to any (consistent) spatially variable boundary data}, are bounded by the energy produced in the composite by uniform tractions (for a common average strain) and/or constant velocities (for a common average momentum) boundary conditions.

There has been considerable research in the field of variational principles for wave equations (\cite{willis1981avariational,willis1981bvariational,willis1984variational,cherkaev1994variational,altay2004fundamental,milton2009minimization,milton2010minimum} and references therein).
In the present paper we show that the total elastodynamic energy, and the total complementary elastodynamic 
energy of the equivalent solid, when regarded as functionals of the eigenstress (eigenstrain), and eigenmomentum (eigenvelocity), are stationary for the exact eigenstress (eigenstrain), and eigenmomentum (eigenvelocity). 
These are the dynamic equivalents of the Hashin-Shtrikman variational principles and are consistent with results in published literature (\cite{willis1981bvariational,willis1984variational,milton2010minimum}). 
In addition, we develop strict (and computable) bounds for these energies that apply to \emph{any spatially variable} (consistent) boundary data. 

\section{Problem Definition and Introductory Results}

Consider the dynamics of a general heterogeneous solid which
consists of various elastic phases. There is no restriction
on the number, geometry, material, or orientation of each constituting
microphase. Consider an arbitrary finite sample of volume $\Omega$ of boundary $\partial\Omega$. The sample is characterized by spatially varying real-valued and positive-definite stiffness tensor, $\mathbf{C(x)}$, with rectangular Cartesian components $C_{ijkl}=C_{jikl}=C_{ijlk}=C_{klij}, (i, j, k, l =1, 2, 3)$, and real-valued positive density, $\rho\mathbf{(x)}$. 
The corresponding constitutive relations are
\begin{equation}
\begin{array}{l}
\displaystyle\boldsymbol{\varepsilon}= \mathbf{D}:\boldsymbol{\sigma};\quad
\displaystyle \dot{\mathbf{u}}={\nu}\mathbf{p},
\end{array} 
\label{Strain_Stress}
\end{equation}
\begin{equation}
\begin{array}{l}
\displaystyle\boldsymbol{\sigma}= \mathbf{C}:\boldsymbol{\varepsilon};\quad
\displaystyle \mathbf{p}={\rho}\dot{\mathbf{u}},
\end{array} 
\label{Stress_Strain}
\end{equation}
or, in components form,
\begin{equation}
\begin{array}{l}
\displaystyle \varepsilon_{ij}=D_{ijkl}\sigma_{kl};\quad
\dot{u}_i={\nu}p_i,\\
\end{array} 
\label{Strain_Stress_Comp}
\end{equation}
\begin{equation}
\begin{array}{l}
\displaystyle \sigma_{ij}=C_{ijkl}\varepsilon_{kl};\quad
p_i=\rho\dot{u}_i,\\
\end{array} 
\label{Stress_Strain_Comp}
\end{equation}
where the repeated indices are summed, ${\nu}(\mathbf{x})=\rho^{-1}$ is the specific volume, and $\mathbf{D(\mathbf{x})}=\mathbf{C}^{-1}$ is the compliance tensor. A given dynamic boundary data (tractions, velocities, or mixed) produces in $\Omega$ variable fields that we collectively represent as:
\begin{equation}\label{EPeriodic}
{\hat{\mathbf{Q}}(\mathbf{x},t)}=\mathrm{Re}\left[\mathbf{Q}(\mathbf{x})\exp^{-i\omega t}\right],
\end{equation}
where the frequency $\omega$ is real-valued and fixed, $\hat{\mathbf{Q}}$ represents any of the field variables, stress $\hat{\boldsymbol{\sigma}}$, strain $\hat{\boldsymbol{\varepsilon}}$, momentum $\hat{\mathbf{p}}$ or velocity $\hat{\dot{\mathbf{\mathbf{u}}}}$, with respective components, $ \hat{\sigma}_{ij}$, $\hat{\varepsilon}_{ij}$, $\hat{p}_j$ or $\hat{\dot{u}}_i$, whereas $\mathbf{Q}$ represents the corresponding spatially variable part. Define the volume average of a typical field quantity $ \mathbf{Q}(\mathbf{x}) $ and its deviation from the volume average, $ \mathbf{Q}^d(\mathbf{x}) $, respectively by
\begin{equation}
\langle\mathbf{Q}\rangle=\frac{1}{\Omega}\int_{\Omega}\mathbf{Q}(\mathbf{x})dV;\quad\mathbf{Q}^d(\mathbf{x})=\mathbf{Q}(\mathbf{x})-\langle\mathbf{Q}\rangle.
\end{equation}
\label{Average_Disturbance}
The local conservation and kinematic relations are
\begin{equation}
\begin{array}{l}
\displaystyle{\nabla}\cdot\boldsymbol{\sigma}=-i\omega\mathbf{p};\quad
\displaystyle ({\nabla}\otimes\dot{\mathbf{u}})_{sym}=-i\omega\boldsymbol{{\varepsilon}},\quad{{in}\ \Omega}.\\
\end{array} 
\label{EFieldEqn}
\end{equation}
Here, the operator ${\nabla}$ stands for $\frac{\partial}{\partial x_{j}}, j=1, 2, 3$.
The boundary data may be general where tractions or velocities may be prescribed on parts of $\partial\Omega$.  To be specific consider
\begin{equation}
\begin{array}{l}
\mathbf{n}\cdot\boldsymbol{\sigma}=\mathbf{t}^B,\quad on\ \partial{\Omega}_T;\quad
\dot{\mathbf{u}}=\dot{\mathbf{u}}^B,\quad on\ \partial{\Omega}_U,\\
\end{array} 
\label{B-Cs}
\end{equation}
where $\mathbf{t}^B(\mathbf{x})$ and $\dot{\mathbf{u}}^B(\mathbf{x})$ are the tractions and velocities that are prescribed on $\partial{\Omega}_T$ and $\partial{\Omega}_U$, respectively, with $\partial{\Omega}= \partial{\Omega}_T+\partial{\Omega}_U$, and $\mathbf{n}$ with components $n_{i}$ is the exterior unit normal on $\partial{\Omega}$.  To simplify notation in Eqs. (\ref{EFieldEqn}, \ref{B-Cs}) and in what follows the dependence of the field variables on $\mathbf{x}$ is not explicitly displayed.

The field equations (\ref{EFieldEqn}) hold for any composition of $\Omega$.   They  imply certain useful identities. Together with constitutive relations (\ref{Strain_Stress}, \ref{Stress_Strain}), they also imply two general \emph{energy bounds} that are essential for establishing strict and computable bounds for the overall effective dynamic constitutive parameters of a given composite, valid for \emph{any} (consistent) boundary conditions. These results are summarized below.

\subsection{Material independent identities}

Define the scalar inner products between two complex-valued vectors, $\mathbf{S}^1\mathbf{(x)}$ and $\mathbf{S}^2\mathbf{(x)}$, and two complex-valued second-order symmetric tensors, $\mathbf{T}^1\mathbf{(x)}$ and $\mathbf{T}^2\mathbf{(x)}$ as follows:
\begin{equation}
\begin{array}{l}
\displaystyle\langle\mathbf{S}^1\cdot\mathbf{S}^2\rangle=\frac{1}{\Omega}\int_{\Omega}S^1_i(\mathbf{x})\left[S^2_i(\mathbf{x})\right]^*dV_\mathbf{x};\\
\displaystyle\langle\mathbf{T}^1:\mathbf{T}^2\rangle=\frac{1}{\Omega}\int_{\Omega}T^1_{ij}(\mathbf{x})\left[T^2_{ij}(\mathbf{x})\right]^*dV_\mathbf{x}.
\end{array} 
\end{equation}
The inner product is a complex-valued scalar and the symbol * denotes a complex conjugate. 
Now, Eqs. (\ref{EFieldEqn}) directly yield the following material-independent identities:
\begin{equation}
\langle\varepsilon_{ij}\rangle=\frac{1}{\Omega}\int_{\partial\Omega}\frac{1}{2}\left[n_iu_j+n_ju_i\right]dS;
\end{equation}
\begin{equation}
\langle\sigma_{ij}\rangle=\frac{1}{\Omega}\int_{\partial\Omega}\frac{1}{2}\left[t_ix_j+t_jx_i\right]dS-\frac{1}{2}\langle\dot{p}_ix_j+\dot{p}_jx_i\rangle;
\end{equation}
\begin{equation}
\langle\sigma_{ij,j}\rangle=\langle\dot{p}_i\rangle=\frac{1}{\Omega}\int_{\partial\Omega}t_idS;
\end{equation}
\begin{equation}\label{ModifiedHill}
\{\langle\boldsymbol{\sigma}:\boldsymbol{\varepsilon}\rangle+\langle \mathbf{p}\cdot\dot{\mathbf{u}}\rangle\}-\{\langle\boldsymbol{\sigma}\rangle:\langle\boldsymbol{\varepsilon}\rangle^*+\langle \mathbf{p}\rangle\cdot\langle\dot{\mathbf{u}}\rangle^*\}
=\frac{1}{\Omega}\int_{\partial\Omega}(\mathbf{t}^B-\mathbf{n}\cdot\langle\boldsymbol{\sigma}\rangle)\cdot(\mathbf{u}^B-\langle \mathbf{u}\rangle)^*dS.
\end{equation}
The final identity (\ref{ModifiedHill}) is the dynamic equivalent of Hill's (static-case) identity.  We shall refer to  (\ref{ModifiedHill}) as the \emph{generalized elastodynamic Hill identity}.  
 
\subsection{General energy bounds for elastic composites}

Consider an elastic composite $\Omega$ consisting of any set of elastic constituents.  There are two universal theorems that provide bounds for the total \emph{elastostatic strain energy} and the total \emph{complementary elastostatic energy} of $\Omega$ subjected to any consistent boundary data; see \cite{nemat1995universal,nemat1999micromechanics}. These are:

\begin{itemize}
\item \emph{Theorem I}: Among all consistent boundary data that produce in a given composite the same overall average strain, $\langle\boldsymbol{\varepsilon}\rangle$, the uniform-tractions boundary data render the total elastostatic energy an absolute minimum.
\item \emph{Theorem II}: Among all consistent boundary data that produce in a given composite the same overall average stress, $\langle\boldsymbol{\sigma}\rangle$, the linear-displacements boundary data render the total complementary elastostatic energy an absolute minimum.

\end{itemize}

These theorems have recently been generalized to \emph{elastodynamic} cases by \cite{srivastava2011universal}.  
Define the total \emph{elastodynamic}  (\emph{elastic strain  plus the kinetic }) energy, and the total \emph{complementary elastodynamic} (\emph{complementary elastic strain plus kinetic})  energy of $\Omega$, respectively as
\begin{equation}\label{Strain_Energy}
\Pi(\mathbf{C},\rho)=\frac{1}{2}\{\langle\varepsilon_{ij}C_{ijkl}\varepsilon_{kl}^*\rangle+\langle\dot{u}_i\rho{\dot{u}_i}^*\rangle\};
\end{equation}
\begin{equation}\label{Complementary_Energy}
\Pi^c(\mathbf{D},\nu)=\frac{1}{2}\{\langle\sigma_{ij}D_{ijkl}\sigma_{kl}^*\rangle+\langle{p}_i\nu{{p}_i}^*\rangle\},
\end{equation}
where the arguments $\mathbf{C}$, $\mathbf{D}$, $\rho$, and $\nu$ 
indicate the corresponding dependence on the (spatially variable) constitutive parameters.  
For any consistent and nonzero boundary data, these energy quantities are real-valued and positive. 
The dynamic versions (for a fixed frequency, $\omega$) of theorems \emph{I} and \emph{II} are as follows:

\begin{itemize}
\item \emph{Theorem $D_1$}: At a fixed frequency $\omega$, among all consistent boundary data that produce in a given composite $\Omega$, the same overall average strain $\langle\boldsymbol{\varepsilon}\rangle$, the uniform boundary tractions render the total \emph{elastodynamic energy} $\Pi$, an absolute minimum.
\item \emph{Theorem $D_2$}: At a fixed frequency $\omega$, among all consistent boundary data that produce in a given composite $\Omega$, the same overall average momentum $\langle \mathbf{p}\rangle$, the constant boundary velocities render the total \emph{complementary elastodynamic energy} $\Pi^c$,  an absolute minimum.

\end{itemize}

Theorems $D_1$ and $D_2$ play key roles in establishing bounds for the overall constitutive parameters of the composite $\Omega$.  
The boundary data for the composite are completely general and arbitrary. 
They must of course be self-consistent and, for $D_1$, they must produce a common volume average strain tensor, while for $D_2$, they must produce a common volume average linear momentum vector.

When these general boundary data are restricted such that the surface integral in the right-hand side of (\ref{ModifiedHill}) vanishes, then the total energy of $\Omega$ can be calculated using the averaged field quantities.  That is, in such a case, the \emph{generalized elastodynamic Hill identity} (\ref{ModifiedHill}) yields the following identity:
\begin{equation}\label{Special_Hill}
\langle\boldsymbol{\sigma}:\boldsymbol{\varepsilon}\rangle+\langle \mathbf{p}\cdot\dot{\mathbf{u}}\rangle=\langle\boldsymbol{\sigma}\rangle:\langle\boldsymbol{\varepsilon}\rangle^*+\langle \mathbf{p}\rangle\cdot\langle\dot{\mathbf{u}}\rangle^*.
\end{equation}
We shall refer to  (\ref{Special_Hill}) as the \emph{generalized elastodynamic Hill condition}. Define the disturbance tractions and the disturbance velocities as
\begin{equation}
\mathbf{t}^d=\mathbf{t}^B -\mathbf{n}\cdot{\langle\boldsymbol{\sigma}\rangle};
\quad\dot{\mathbf{u}}^d=\dot{\mathbf{u}}^B-\langle\dot{\mathbf{u}}\rangle.
\end{equation}
Then the equality (\ref{Special_Hill}) holds for all boundary data for which
\begin{equation}\label{BD_Zero}
\int_{\partial\Omega}\mathbf{t}^d\cdot\dot{\mathbf{u}}^{d*}dS=\mathbf{0}.
\end{equation}
For this class of boundary data, the total energy of the finite composite can be computed using the averaged  stress, strain, momentum, and velocity. This class of boundary data includes special cases of uniform tractions and uniform velocities boundary data which are of special importance, as discussed later on. Theorems $D_1$ and $D_2$ remain valid whether or not (\ref{BD_Zero}) holds.  
However, for the class of boundary data which satisfy (\ref{BD_Zero}),  
the generalized dynamic Hashin-Shtrikman bounds can be expressed directly in terms of the effective dynamic stiffness (compliance) and mass-density (specific mass) tensors of the composite.  
\emph{In what follows, we focus on boundary data that do satisfy (\ref{BD_Zero})}.

\subsection{Effective elastodynamic constitutive relations}

Effective dynamic constitutive parameters are defined by relating the volume averages of the field variables. These field variables are functions of the boundary data. 
Therefore, the resulting constitutive coefficients depend upon the boundary conditions on $\partial\Omega$. 
The general form of the overall constitutive relations is given by
(\cite{milton2007modifications,willis2009exact,nematnasser2011overall,willis2011effective,srivastava2011overall}),
\begin{equation}\label{DynamicRelation1}
\begin{array}{l}
\displaystyle \langle\boldsymbol{\sigma}\rangle=\bar{\mathbf{C}}:\langle\boldsymbol{\varepsilon}\rangle+\bar{\mathbf{S}}\cdot\langle\dot{\mathbf{u}}\rangle;\\
\displaystyle \langle\mathbf{p}\rangle=\bar{\mathbf{S}}^\dagger:\langle\boldsymbol{\varepsilon}\rangle+\bar{\boldsymbol{\rho}}\cdot\langle\dot{\mathbf{u}}\rangle.
\end{array} 
\end{equation}
The effective constitutive parameters reflect the non-local spatial microstructure. 
They may be complex-valued even if the composite is non-dissipative, the imaginary parts resulting from the asymmetries of, e.g., the unit cell of a periodic composite.  
They are functions of the frequency, $\omega$, and in the case of elastic waves, they also depend on the 
wavevector, $\mathbf{q}$.
$\bar{\mathbf{C}}$ is the fourth-order effective elasticity tensor which has minor symmetries, 
$\bar{C}_{ijkl}=\bar{C}_{jikl}=\bar{C}_{ijlk}$. 
It does not have the major symmetry associated with the elasticity or the compliance tensor but has a hermitian relationship over the major transformation, $\bar{C}_{ijkl}=[\bar{C}_{klij}]^*$. 
Effective density is a second-order tensor with a hermitian relationship over the transformation of its indices, 
$\bar{\rho}_{ij}=[\bar{\rho}_{ji}]^*$, and $\bar{\mathbf{S}}$ 
is the third-order coupling tensor with a hermitian transpose given by 
$\bar{S}_{kij}^\dagger=\bar{S}_{ijk}^*$. 

Since the cases under consideration satisfy (\ref{BD_Zero}), 
the total elastodynamic energy stored within the domain follows directly from Eqs. (\ref{Special_Hill}, \ref{DynamicRelation1}):
\begin{equation}\label{Energy_1}
\begin{array}{c}
\displaystyle \Pi(\bar{\mathbf{C}},\bar{\boldsymbol{\rho}})=\frac{1}{2}\left[\langle\boldsymbol{\varepsilon}\rangle^*:\langle\boldsymbol{\sigma}\rangle+\langle\dot{\mathbf{u}}\rangle^*\cdot\langle\mathbf{p}\rangle\right]
\displaystyle=\frac{1}{2}\{\langle\boldsymbol{\varepsilon}\rangle^*:\bar{\mathbf{C}}:\langle\boldsymbol{\varepsilon}\rangle+\langle\dot{\mathbf{u}}\rangle^*\cdot\bar{\boldsymbol{\rho}}\cdot\langle\dot{\mathbf{u}}\rangle\\
\displaystyle \quad+\langle\boldsymbol{\varepsilon}\rangle^*:\bar{\mathbf{S}}\cdot\langle\dot{\mathbf{u}}\rangle+\langle\dot{\mathbf{u}}\rangle^*\cdot\bar{\mathbf{S}}^\dagger:\langle\boldsymbol{\varepsilon}\rangle\}.
\end{array} 
\end{equation}
Given the structure of the constitutive tensors, it can be seen that the total elastodynamic energy as expressed above, is strictly real-valued. 
Alternatively, the averaged constitutive relations can be expressed as
\begin{equation}\label{DynamicRelation2}
\begin{array}{l}
\displaystyle \langle\boldsymbol{\varepsilon}\rangle=\bar{\mathbf{D}}:\langle\boldsymbol{\sigma}\rangle+\bar{\mathbf{R}}\cdot\langle\mathbf{p}\rangle;\\
\displaystyle \langle\dot{\mathbf{u}}\rangle=\bar{\mathbf{R}}^\dagger:\langle\boldsymbol{\sigma}\rangle+\bar{\boldsymbol{\nu}}\cdot\langle\mathbf{p}\rangle.
\end{array} 
\end{equation}
in which case the total complementary elastodynamic energy stored within $\Omega$ becomes
\begin{equation}\label{Energy_2}
\begin{array}{c}
\displaystyle \Pi^c(\bar{\mathbf{D}},\bar{\boldsymbol{\nu}})=\frac{1}{2}\left[\langle\boldsymbol{\sigma}\rangle^*:\langle\boldsymbol{\varepsilon}\rangle+\langle\mathbf{p}\rangle^*\cdot\langle\dot{\mathbf{u}}\rangle\right]=\frac{1}{2}\{\langle\boldsymbol{\sigma}\rangle^*:\bar{\mathbf{D}}:\langle\boldsymbol{\sigma}\rangle+\langle\mathbf{p}\rangle^*\cdot\bar{\boldsymbol{\nu}}\cdot\langle\mathbf{p}\rangle\\
\displaystyle \quad+\langle\boldsymbol{\sigma}\rangle^*:\bar{\mathbf{R}}\cdot\langle\mathbf{p}\rangle+\langle\mathbf{p}\rangle^*\cdot\bar{\mathbf{R}}^\dagger:\langle\boldsymbol{\sigma}\rangle\}.
\end{array} 
\end{equation}
which is also real-valued.

%
%

\section{Equivalent Homogeneous Solid}

We replace the heterogeneous $\Omega$ by a geometrically identical but materially homogeneous one having (conveniently selected) uniform density $\rho^0$ and elasticity $\mathbf{C}^0$.  
Denote this homogeneous solid by $\Omega^*$. 
Then we introduce in $\Omega^*$, \emph{eigenstress} $\mathbf{\Sigma}(\mathbf{x})$ and \emph{eigenmomentum} ${\mathbf{P}}(\mathbf{x})$ such that the stress $\boldsymbol{\sigma}(\mathbf{x})$ and momentum ${\mathbf{p}}(\mathbf{x})$ at every point within the homogenized  $\Omega^*$ are exactly the same as they are in the original heterogeneous $\Omega$.
 
Alternatively, we can use uniform specific volume $\nu^0$ and compliance $\mathbf{D}^0$ for the homogeneous $\Omega^*$, and introduce \emph{eigenstrain} $\mathbf{E}(\mathbf{x})$ and \emph{eigenvelocity} $\dot{\mathbf{U}}(\mathbf{x})$ such that the strain $\boldsymbol{\varepsilon}(\mathbf{x})$ and velocity $\dot{\mathbf{u}}(\mathbf{x})$ at every point within the homogenized  $\Omega^*$  are exactly the same as they are in the original heterogeneous $\Omega$. In general the two homogenization methods may not be equivalent.  But, for the class of boundary data which satisfy the generalized elastodynamic Hill condition they are equivalent.

 
\section{Homogenization Using Eigenstress and Eigenmomentum and Associated Bounds} 

Consider the homogeneous $\Omega^*$, subjected to the prescribed boundary conditions which must satisfy (\ref{BD_Zero}).  
The complex-valued eigenstress and eigenmomentum, 
${\mathbf{\Sigma}(\mathbf{x})}$ and ${\mathbf{P}}(\mathbf{x})$, 
must be such that the following \emph{consistency conditions} hold at every point within $\Omega^*$:
\begin{equation}
\boldsymbol{\sigma}=\mathbf{C}:\boldsymbol{\varepsilon}=\mathbf{C}^0:\boldsymbol{\varepsilon}+\mathbf{\Sigma};
\quad\displaystyle {\mathbf{p}}={\rho}\dot{\mathbf{u}}={\rho}^0\dot{\mathbf{u}}+{\mathbf{P}},\\
\label{Eeigenstuff}
\end{equation}
or, in components form,
\begin{equation}\label{Eeigenstuff_comp}
\sigma_{ij}=C_{ijkl}{\varepsilon}_{kl}=C^0_{ijkl}\sigma_{kl}+\Sigma_{ij};
\quad{p}_i={\rho}\dot{{u}}_i={\rho}^0\dot{u}_i+{P}_i.
\end{equation}
For eigenstress and eigenmomentum, $\mathbf{{\Sigma}}$ and ${\mathbf{P}}$,  
the corresponding disturbance fields,  
\begin{equation}\label{Disturb_Fields1}
\boldsymbol{\varepsilon}^d(\mathbf{x})=\boldsymbol{\varepsilon}(\mathbf{x})-\langle\boldsymbol{\varepsilon}\rangle;\
\quad\dot{\mathbf{u}}^d(\mathbf{x})=\dot{\mathbf{u}}(\mathbf{x})-\langle\dot{\mathbf{u}}\rangle, 
\end{equation}
can be expressed in terms of integral operators:
\begin{equation}\label{IntegralOp1}
\begin{array}{c}
\displaystyle \boldsymbol{\varepsilon}^d\mathbf{(x)}=-\left[\boldsymbol{\Gamma}^{(\boldsymbol{\varepsilon}\mathbf{P})}(\mathbf{x};\mathbf{P})+\boldsymbol{\Gamma}^{(\boldsymbol{\varepsilon}\mathbf{\Sigma})}(\mathbf{x};\mathbf{\Sigma})\right],\\

\displaystyle \dot{\mathbf{u}}^d\mathbf{(x)}=-\left[\boldsymbol{\Gamma}^{(\dot{\mathbf{u}}\mathbf{P})}(\mathbf{x};\mathbf{P})+\boldsymbol{\Gamma}^{(\dot{\mathbf{u}}\mathbf{\Sigma})}(\mathbf{x};\mathbf{\Sigma})\right].\\

\end{array} 
\end{equation}
It is emphasized that the integral operators in the right-hand side of the above expressions 
depend on the geometry and the boundary conditions of $\partial{\Omega}^*$.  
In \ref{A}, exact expressions for these integral operators are provided 
for a periodic or a parallelepiped-shaped finite composite using a Fourier series representation. 
In what follows it will prove convenient to rewrite the above integral operators as: 
\begin{equation}\label{Gamma_Operators}
\begin{array}{l}
\boldsymbol{\Gamma}^{(\boldsymbol{\varepsilon}\mathbf{P})}(\mathbf{x};\hat{\mathbf{P}})\equiv\displaystyle\hat{\mathbf{P}}\cdot \boldsymbol{\Gamma}^{(\boldsymbol{\varepsilon}\mathbf{P})};\quad

\boldsymbol{\Gamma}^{(\boldsymbol{\varepsilon}\mathbf{\Sigma})}(\mathbf{x};\hat{\mathbf{\Sigma}})\equiv\hat{\mathbf{\Sigma}}:\boldsymbol{\Gamma}^{(\boldsymbol{\varepsilon}\mathbf{\Sigma})},\\

\boldsymbol{\Gamma}^{(\dot{\mathbf{u}}\mathbf{P})}(\mathbf{x};\hat{\mathbf{P}})\equiv\displaystyle\hat{\mathbf{P}}\cdot \boldsymbol{\Gamma}^{(\dot{\mathbf{u}}\mathbf{P})};\quad

\boldsymbol{\Gamma}^{(\dot{\mathbf{u}}\mathbf{\Sigma})}(\mathbf{x};\hat{\mathbf{\Sigma}})\equiv\hat{\mathbf{\Sigma}}:\boldsymbol{\Gamma}^{(\dot{\mathbf{u}}\mathbf{\Sigma})},
\end{array}
\end{equation}
keeping in mind that these operations are different from simple tensorial contractions, 
and that the operators do depend on the boundary data.
Now the consistency conditions can be written in terms of the integral operators as follows:
\begin{equation}\label{Consistency3}
\mathbf{\Sigma}:\left[\mathbf{C}-\mathbf{C}^0\right]^{-1}+\mathbf{P}\cdot\boldsymbol{\Gamma}^{(\boldsymbol{\varepsilon}\mathbf{P})}+\mathbf{\Sigma}:\boldsymbol{\Gamma}^{(\boldsymbol{\varepsilon}\mathbf{\Sigma})}-\langle\boldsymbol{\varepsilon}\rangle=\mathbf{0},
\end{equation}
\begin{equation}\label{Consistency4}
\mathbf{P}\left[{\rho}-{\rho}^0\right]^{-1}+\mathbf{P}\cdot\boldsymbol{\Gamma}^{(\dot{\mathbf{u}}\mathbf{P})}+\mathbf{\Sigma}:\boldsymbol{\Gamma}^{(\dot{\mathbf{u}}\mathbf{\Sigma})}-\langle\dot{\mathbf{u}}\rangle=\mathbf{0}.
\end{equation}
These are two coupled integral equations that define the homogenizing eigenfields, $\mathbf{\Sigma}$ and $\mathbf{P}$, 
for given $\langle{\boldsymbol{\varepsilon}}\rangle$ and $\langle{\dot{\mathbf{u}}}\rangle$. It can be shown that the quadratic functional given by
\begin{equation}\label{Functional_1}
\begin{array}{c}
\displaystyle \mathcal{F}(\hat{\mathbf{\Sigma}},\hat{\mathbf{P}})\equiv\\

\displaystyle \frac{1}{2}\left[\langle\hat{\mathbf{\Sigma}}:\
\{(\mathbf{C}-\mathbf{C}^0)^{-1}+\
\boldsymbol{\Gamma}^{(\boldsymbol{\varepsilon}{\mathbf{\Sigma}})}\}:\hat{\mathbf{\Sigma}}\rangle+\langle\hat{\mathbf{P}}\cdot\{(\rho-\rho^0)^{-1}+\
\boldsymbol{\Gamma}^{(\dot{\mathbf{u}}{\mathbf{P}})}\}\cdot\hat{\mathbf{P}}\rangle\right.\\

\displaystyle \left.+\{\langle\hat{\mathbf{P}}\cdot\boldsymbol{\Gamma}^{(\boldsymbol{\varepsilon}{\mathbf{P}})}:\hat{\mathbf{\Sigma}}\rangle+\langle\hat{\mathbf{\Sigma}}:\boldsymbol{\Gamma}^{(\dot{\mathbf{u}}{\mathbf{\Sigma}})}\cdot\hat{\mathbf{P}}\rangle\}\right],

\end{array} 
\end{equation}
is real-valued and that, for eigenfields which result in field variables satisfying the generalized Hill condition, it is (\ref{B}):
\begin{itemize}
\item positive when $(\mathbf{C}-\mathbf{C}^0)$ is positive-definite 
and $(\rho-\rho^0)$ is positive;
\item negative when $(\mathbf{C}-\mathbf{C}^0)$ is negative-definite and $(\rho-\rho^0)$ is negative.
\end{itemize}
Additionally, for the exact eigenfields, $\mathbf{\Sigma}(\mathbf{x})$ and $\mathbf{P}(\mathbf{x})$, that satisfy the consistency conditions (\ref{Consistency3}, \ref{Consistency4}):
\begin{equation}\label{Functional_2}
\mathcal{F}(\mathbf{\Sigma},\mathbf{P})=\
\frac{1}{4}\left[\{\langle\boldsymbol{\varepsilon}\rangle:\langle{\mathbf{\Sigma}}\rangle^*+\langle\boldsymbol{\varepsilon}\rangle^*:\langle{\mathbf{\Sigma}}\rangle\}+\{\langle\dot{\mathbf{u}}\rangle\cdot\langle\mathbf{P}\rangle^*+\langle\dot{\mathbf{u}}\rangle^*\cdot\langle\mathbf{P}\rangle\}\right].
\end{equation}
Now averaging (\ref{Eeigenstuff}) over $\Omega$, we have
\begin{equation}\label{Average_Const}
\begin{array}{l}
\langle\boldsymbol{\sigma}\rangle=\
\mathbf{C}^0:\langle\boldsymbol{\varepsilon}\rangle+\
\langle\mathbf{\Sigma}\rangle,\\

\langle\mathbf{p}\rangle=\
\rho^0\langle\dot{\mathbf{u}}\rangle+\
\langle\mathbf{P}\rangle.
\end{array} 
\end{equation}
In view of Eqs. (\ref{Functional_2}, \ref{Average_Const}) and using (\ref{DynamicRelation1}), it now follows that
\begin{equation}\label{Functional_3}
\begin{array}{c}
\displaystyle \mathcal{F}(\mathbf{\Sigma},\mathbf{P})=\
\displaystyle \frac{1}{2}\left[\
\langle\boldsymbol{\varepsilon}\rangle^*:\
(\bar{\mathbf{C}}-\mathbf{C}^0):\
\langle\boldsymbol{\varepsilon}\rangle+\
\langle\dot{\mathbf{u}}\rangle^*\cdot(\bar{\boldsymbol{\rho}}-\rho^0\mathbf{I})\
\cdot{\langle\dot{\mathbf{u}}\rangle}\right.\\

\left.+\{\langle\boldsymbol{\varepsilon}\rangle^*:\
\bar{\mathbf{S}}\cdot\
\langle\dot{\mathbf{u}}\rangle+\
\langle\dot{\mathbf{u}}\rangle^*\cdot\
\bar{\mathbf{S}}^\dag:\
\langle\boldsymbol{\varepsilon}\rangle\}\right]\\

\equiv \Pi(\{\bar{\mathbf{C}}-\mathbf{C}^0\},\{\bar{\boldsymbol{\rho}}-\rho^0\mathbf{I}\}).
\end{array} 
\end{equation}
Eqs. (\ref{Functional_2}-\ref{Functional_3}) hold only for the \emph{averaged values of the exact homogenizing eigenfields}, $\langle{\mathbf{\Sigma}}\rangle$ and $\langle{\mathbf{P}}\rangle$. 
As is seen, $\mathcal{F}(\mathbf{\Sigma},\mathbf{P})$ is 
the elastodynamic energy of the homogenized $\Omega^*$ 
when its effective dynamic elasticity is  $(\bar{\mathbf{C}}-\mathbf{C}^0)$, 
its effective dynamic density is $(\bar{\boldsymbol{\rho}}-\rho^0\mathbf{I})$, 
and it is supporting the average strain $\langle\boldsymbol{\varepsilon}\rangle$ and average velocity $\langle\dot{\mathbf{u}}\rangle$ at a fixed frequency $\omega$.

\subsection{A Hashin-Shtrikman-type variational principle}

Based on the total elastodynamic energy of $\Omega$ and for any arbitrary eigenstress, $\hat{\mathbf{\Sigma}}$, and eigenmomentum, $\hat{\mathbf{P}}$, fields prescribed in the equivalent homogeneous $\Omega^*$, define a Hashin-Shtrikman-type functional as follows:
\begin{equation}\label{S_H_Functional_1}
\begin{array}{c}
\displaystyle \mathcal{J}(\hat{\mathbf{\Sigma}},\hat{\mathbf{P}};\langle\boldsymbol{\varepsilon}\rangle,\langle\dot{\mathbf{u}}\rangle)=\\

\displaystyle \mathcal{F}(\hat{\mathbf{\Sigma}},\hat{\mathbf{P}})-\

\displaystyle \frac{1}{2}\left[\langle\boldsymbol{\varepsilon}\rangle:\langle\hat{\mathbf{\Sigma}}\rangle^*+\
\langle \hat{\mathbf{\Sigma}}\rangle:\
\langle\boldsymbol{\varepsilon}\rangle^*+\ 
 \langle\dot{\mathbf{u}}\rangle\cdot\langle\hat{\mathbf{P}}\rangle^*+\
 \langle \hat{\mathbf{P}}\rangle\cdot\langle\dot{\mathbf{u}}\rangle^*\right],

\end{array} 
\end{equation}
where $\langle\boldsymbol{\varepsilon}\rangle$ and  $\langle\dot{\mathbf{u}}\rangle$ are the volume averages of the strain and velocity in $\Omega$, while the field variables, $\hat{\mathbf{\Sigma}}$ and $\hat{\mathbf{P}}$, are functions subject to arbitrary variations. Using the symmetries of the tensors shown in \ref{A} it can be shown that the variations of the $\mathcal{J}$ functional with respect to the independent variations of the field variables $\hat{\mathbf{\Sigma}}$ and $\hat{\mathbf{P}}$ are given by
\begin{equation}
\begin{array}{l}
\displaystyle
[\mathcal{J}(\hat{\mathbf{\Sigma}},\hat{\mathbf{P}};\langle\boldsymbol{\varepsilon}\rangle,\langle\dot{\mathbf{u}}\rangle)]_{\delta\hat{\mathbf{\Sigma}}}=\langle\left[\hat{\mathbf{\Sigma}}:(\mathbf{C}-\mathbf{C}^0)^{-1}+\hat{\mathbf{P}}\cdot\mathbf{\Gamma}^{(\boldsymbol{\varepsilon}{\mathbf{P}})}+\hat{\mathbf{\Sigma}}:\mathbf{\Gamma}^{(\boldsymbol{\varepsilon}{\mathbf{\Sigma}})}-\langle\boldsymbol{\varepsilon}\rangle\right]:\delta\hat{\mathbf{\Sigma}}\rangle;\\\\

\displaystyle
[\mathcal{J}(\hat{\mathbf{\Sigma}},\hat{\mathbf{P}};\langle\boldsymbol{\varepsilon}\rangle,\langle\dot{\mathbf{u}}\rangle)]_{\delta\hat{\mathbf{P}}}=\langle\left[\hat{\mathbf{P}}({\rho}-{\rho}^0)^{-1}+\hat{\mathbf{P}}\cdot\mathbf{\Gamma}^{(\dot{\mathbf{u}}{\mathbf{P}})}+\hat{\mathbf{\Sigma}}:\mathbf{\Gamma}^{(\dot{\mathbf{u}}{\mathbf{\Sigma}})}-\langle\dot{\mathbf{u}}\rangle\right]\cdot\delta\hat{\mathbf{P}}\rangle.
\end{array}
\end{equation} 
It is seen from Eqs. (\ref{Consistency3}, \ref{Consistency4}) that the above variations go to 0 for the \emph{exact} eigenstress and eigenmomentum fields, 
$\mathbf{{\Sigma}}$ and ${\mathbf{P}}$, which produce in the equivalent homogeneous solid the same stress and momentum fields as in the original heterogeneous $\Omega$ subjected to given boundary data. 
Hence, $\mathcal{J}({\mathbf{\Sigma}},{\mathbf{P}};\langle\boldsymbol{\varepsilon}\rangle,\langle\dot{\mathbf{u}}\rangle)=-\mathcal{F}({\mathbf{\Sigma}},{\mathbf{P}})
=-\Pi(\{\bar{\mathbf{C}}-\mathbf{C}^0\},\{\bar{\boldsymbol{\rho}}-\rho^0\mathbf{I}\})$ is the stationary value of Eq. (\ref{S_H_Functional_1}).
 Moreover, the vanishing of the  variations of $\mathcal{J}(\hat{\mathbf{\Sigma}},\hat{\mathbf{P}};\langle\boldsymbol{\varepsilon}\rangle,\langle\dot{\mathbf{u}}\rangle)$ for arbitrary variations of $\hat{\mathbf{\Sigma}}$ and $\hat{\mathbf{P}}$, yields the consistency conditions (\ref{Consistency3} )and (\ref{Consistency4}), respectively.

\subsection{Bounds for the Energy Functional}

In the previous section it was shown that the eigenstress and eigenmomentum fields which satisfy the corresponding consistency conditions render the functional $\mathcal{J}$ stationary. 
Under certain conditions, this stationary value becomes the extremum value of the functional. 
To show this we note that the $\mathcal{J}$ functional can be written as
\begin{equation}\label{IFunctionalMod}
\begin{array}{c}
\displaystyle \mathcal{J}(\hat{\mathbf{\Sigma}},\hat{\mathbf{P}};\langle\boldsymbol{\varepsilon}\rangle,\langle\dot{\mathbf{u}}\rangle)=\

\displaystyle \mathcal{F}(\hat{\mathbf{\Sigma}}-{\mathbf{\Sigma}},\hat{\mathbf{P}}-{\mathbf{P}})-\
\mathcal{F}(\mathbf{{\Sigma}},\mathbf{P}).

\end{array} 
\end{equation}
which attains its stationary value for $\hat{\mathbf{\Sigma}}={\mathbf{\Sigma}},\hat{\mathbf{P}}={\mathbf{P}}$. 
The sets of eigenfields $\{\mathbf{\Sigma},\mathbf{P}\}$ and $\{\hat{\mathbf{\Sigma}},\hat{\mathbf{P}}\}$ produce field variables which satisfy the generalized elastodynamic Hill condition (\ref{BD_Zero}). 
Hence, the eigenfields $\{(\hat{\mathbf{\Sigma}}-{\mathbf{\Sigma}}), (\hat{\mathbf{P}}-{\mathbf{P}})\}$ also satisfy the generalized elastodynamic Hill condition. Finally as shown in \ref{B}, for arbitrary eigenfields $\hat{\mathbf{\Sigma}}(\mathbf{x}),\hat{\mathbf{P}}(\mathbf{x})$:

\begin{itemize}
\item If $(\mathbf{C}-\mathbf{C}^0)$ is \emph{negative-definite} and 
$(\rho-\rho^0)$ is \emph{negative}, then 
$\mathcal{J}({\mathbf{\Sigma}},{\mathbf{P}};\langle\boldsymbol{\varepsilon}\rangle,\langle\dot{\mathbf{u}}\rangle) =-\mathcal{F}(\mathbf{{\Sigma}},\mathbf{P})$ is the \emph{maximum} value of $\mathcal{J}(\hat{\mathbf{\Sigma}},\hat{\mathbf{P}};\langle\boldsymbol{\varepsilon}\rangle,\langle\dot{\mathbf{u}}\rangle)$;
\item If $(\mathbf{C}-\mathbf{C}^0)$ is \emph{positive-definite} and
$(\rho-\rho^0)$ is \emph{positive}, then
$\mathcal{J}({\mathbf{\Sigma}},{\mathbf{P}};\langle\boldsymbol{\varepsilon}\rangle,\langle\dot{\mathbf{u}}\rangle) =-\mathcal{F}(\mathbf{{\Sigma}},\mathbf{P})$ is the \emph{minimum} value of $\mathcal{J}(\hat{\mathbf{\Sigma}},\hat{\mathbf{P}};\langle\boldsymbol{\varepsilon}\rangle,\langle\dot{\mathbf{u}}\rangle)$. 
\end{itemize}

\subsection{Exact Inequalities}

An exact inequality is obtained and used to bound the effective dynamic properties of $\Omega$. To this end, choose a reference elasticity $\mathbf{C}^0$ such that $\mathbf{C-C}^0$ is negative-semidefinite, and a reference density such that ${\rho}-{\rho}^0$ is negative. Then for any arbitrary strain fields, $\boldsymbol{\varepsilon}$ and $\hat{\boldsymbol{\varepsilon}}$, and any arbitrary velocity fields, $\dot{\mathbf{u}}$ and $\hat{\dot{\mathbf{u}}}$, the following inequality holds:
\begin{equation}\label{Inequality1}
\frac{1}{2}\left[\langle(\boldsymbol{\varepsilon}-\hat{\boldsymbol{\varepsilon}}):(\mathbf{C-C}^0):(\boldsymbol{\varepsilon}-\hat{\boldsymbol{\varepsilon}})\rangle+\langle(\dot{\mathbf{u}}-\hat{\dot{\mathbf{u}}})({\rho}-{\rho}^0)\cdot(\dot{\mathbf{u}}-\hat{\dot{\mathbf{u}}})\rangle\right]\leq 0.
\end{equation}
For arbitrary eigenstress and eigenmomentum fields, $\hat{\mathbf{\Sigma}}$ and $\hat{{\mathbf{P}}}$, and given average strain and velocity fields, $\langle\boldsymbol{\varepsilon}\rangle$ and $\langle\dot{\mathbf{u}}\rangle$, we consider the following strain and velocity fields:
\begin{equation}\label{defineStrainVelocity}
\begin{array}{l}

\displaystyle \boldsymbol{\varepsilon}=\langle\boldsymbol{\varepsilon}\rangle-\mathbf{\Gamma}^{(\boldsymbol{\varepsilon}\hat{\mathbf{\Sigma}}\hat{{\mathbf{P}}})};\quad \hat{\boldsymbol{\varepsilon}}=(\mathbf{C-C}^0)^{-1}:\hat{\mathbf{\Sigma}}\\

\displaystyle \dot{\mathbf{u}}=\langle\dot{\mathbf{u}}\rangle-\mathbf{\Gamma}^{(\dot{\mathbf{u}}\hat{\mathbf{\Sigma}}\hat{{\mathbf{P}}})};\quad \hat{\dot{\mathbf{u}}}=({\rho}-{\rho}^0)^{-1}:\hat{\mathbf{P}}\\

\end{array} 
\end{equation}
where $-\mathbf{\Gamma}^{(\boldsymbol{\varepsilon}\hat{\mathbf{\Sigma}}\hat{{\mathbf{P}}})}=-\{\hat{{\mathbf{P}}}\cdot\boldsymbol{\Gamma}^{(\boldsymbol{\varepsilon}\mathbf{P})}+\hat{\mathbf{\Sigma}}:\boldsymbol{\Gamma}^{(\boldsymbol{\varepsilon}{\mathbf{\Sigma}})}\}$ and 
$-\mathbf{\Gamma}^{(\dot{\mathbf{u}}\hat{\mathbf{\Sigma}}\hat{{\mathbf{P}}})}=-\{\hat{\mathbf{P}}\cdot\boldsymbol{\Gamma}^{(\dot{\mathbf{u}}\mathbf{P})}+\hat{\boldsymbol{\Sigma}}:\boldsymbol{\Gamma}^{(\dot{\mathbf{u}}{\mathbf{\Sigma}})}\}$
are disturbance strain and velocity fields produced in the homogeneuous $\Omega^*$ by the eigenfields,  
$\hat{\mathbf{\Sigma}}$ and $\hat{{\mathbf{P}}}$,
which satisfy the elastodynamic condition (\ref{BD_Zero}). We now substitute (\ref{defineStrainVelocity}) into inequality (\ref{Inequality1}) and after some manipulation (see \ref{B}) obtain, 
\begin{equation}\label{Inequality2}
\begin{array}{c}
\displaystyle \frac{1}{2}\left[\{\langle\boldsymbol{\varepsilon}:\mathbf{C}:\boldsymbol{\varepsilon}\rangle+\langle\dot{\mathbf{u}}\cdot{\rho}\dot{\mathbf{u}}\rangle\}-\{\langle\langle\boldsymbol{\varepsilon}\rangle:\mathbf{C}^0:\langle\boldsymbol{\varepsilon}\rangle\rangle +\langle\langle\dot{\mathbf{u}}\rangle\cdot{\rho}^0\langle\dot{\mathbf{u}}\rangle\rangle\}\right]\\
\displaystyle +\mathcal{J}(\hat{\mathbf{\Sigma}},\hat{\mathbf{P}};\langle\boldsymbol{\varepsilon}\rangle,\langle\dot{\mathbf{u}}\rangle)\leq 0.\\
\end{array} 
\end{equation}
In terms of the total elastodynamic strain energy of the composite, the above inequality becomes,
\begin{equation}\label{PiInequality}
\Pi(\{\mathbf{C}^0-\mathbf{C}\},\{\rho^0-\rho\})\leq\\
\mathcal{J}(\hat{\mathbf{\Sigma}},\hat{\mathbf{P}};\langle\boldsymbol{\varepsilon}\rangle,\langle\dot     {\mathbf{u}}\rangle).
\end{equation}
Since the considered class of boundary data satisfy (\ref{BD_Zero}), inequality (\ref{PiInequality}) 
can be expressed as
\begin{equation}\label{PiInequalityEffective} 
\Pi(\{\mathbf{C}^0-\bar{\mathbf{C}}\},\{\rho^0\mathbf{I}-\bar{\boldsymbol{\rho}}\})\leq\\
\mathcal{J}(\hat{\mathbf{\Sigma}},\hat{\mathbf{P}};\langle\boldsymbol{\varepsilon}\rangle,\langle\dot{\mathbf{u     }}\rangle).
\end{equation}
For arbitrary eigenfields which result in field variables that satisfy the generalized elastodynamic Hill condition on $\partial{\Omega^*}$, 
and provided that the integral operators that define the right-hand side of (\ref{PiInequality}) are given,
the $\mathcal{J}$ functional provides an upper bound for the elastodynamic strain energy of the composite. 
It can be seen that the equality holds if $\hat{\mathbf{\Sigma}}=\mathbf{\Sigma}$ and $\hat{\mathbf{P}}=\mathbf{P}$.

The bounding functional $\mathcal{J}$ in (\ref{PiInequalityEffective}), 
in addition to depending upon the eigenfields, also depends upon the boundary conditions on $\partial{\Omega}$. 
Among all boundary data sets which satisfy the generalized elastodynamic Hill condition (\ref{BD_Zero}), 
and produce a common average strain, that which corresponds to uniform tractions renders the elastodynamic
strain energy an absolute minimum; see Theorem $D_1$. 
Therefore, for the effective dynamic constitutive parameters corresponding to uniform tractions boundary data, 
inequality (\ref{PiInequalityEffective}) provides
computable bounds using the integral operators of 
\emph{any consistent boundary data} that satisfy (\ref{BD_Zero}).
Thus, for given average strain and velocity, $\langle\boldsymbol{\varepsilon}\rangle$ and
$\langle\dot{\mathbf{u}}\rangle$, and a pair of eigenfields,
$\hat{\mathbf{\Sigma}}$ and $\hat{\mathbf{P}}$, we can use functionals
(\ref{Functional_1} and \ref{S_H_Functional_1}) and obtain,

\begin{equation}\label{Inequality_g}
\Pi_{\mathbf{t}}(\{\mathbf{C}^0-\bar{\mathbf{C}}\},\{\rho^0\mathbf{I}-\bar{\boldsymbol{\rho}}\})\leq\\ 
\mathcal{J}_{\mathbf{g}}(\hat{\mathbf{\Sigma}},\hat{\mathbf{P}};\langle\boldsymbol{\varepsilon}\rangle,\langle\dot{\mathbf{u
}}\rangle),
\end{equation}
using any general set of integral operators to compute $\mathcal{J}_{\mathbf{g}}$.
In particular, we can use the operators corresponding to  
periodic boundary data when $\mathbf{\Omega}$ is a parallelepiped or a unit cell of a periodic composite.
These operators are given explicitly in \ref{A}.

\section{Homogenization Using Eigenstrain and Eigenvelocity}

We may homogenize $\Omega$ by introducing in the corresponding homogeneous $\Omega^*$ of uniform specific volume $\nu^0$ and compliance $\mathbf{D}^0$,  the field-variable eigenstrain, $\mathbf{E}$, and eigenvelocity, $\dot{\mathbf{U}}$, such that,  
\begin{equation}
\begin{array}{c}
\displaystyle \boldsymbol{\varepsilon}=\mathbf{D}^0:\boldsymbol{\sigma}+\mathbf{E};\quad \varepsilon_{ij}=D^0_{ijkl}\sigma_{kl}+E_{ij},\\
\displaystyle \dot{\mathbf{u}}={\nu}^0\mathbf{p}+\dot{\mathbf{U}};\quad \dot{u}_i={\nu}^0u_i+\dot{U}_i.\\
\end{array} 
\label{EeigenstuffJ}
\end{equation}
Since at every point within $\Omega$, the strain must be related to the stress by $\boldsymbol{\varepsilon}=\mathbf{D}:\boldsymbol{\sigma}$ and velocity must be related to the momentum by $\dot{\mathbf{u}}={\nu}\mathbf{p}$, it follows that these eigenfields must satisfy the following \emph{consistency conditions}: 
\begin{equation}
\boldsymbol{\varepsilon}=\mathbf{D}(\mathbf{x}):\boldsymbol{\sigma}=\mathbf{D}^0:\boldsymbol{\sigma}+\mathbf{E};
\quad\displaystyle {\dot{\mathbf{u}}}={\nu}(\mathbf{x})\mathbf{p}={\nu}^0\mathbf{p}+\dot{\mathbf{U}}.\\
\end{equation}
In terms of the eigenstrain and eigenvelocity, $\mathbf{{E}}$ and $\dot{\mathbf{U}}$,  the disturbance fields,  $\boldsymbol{\sigma}^d(\mathbf{x})=\boldsymbol{\sigma}(\mathbf{x})-\langle\boldsymbol{\sigma}\rangle$ and $\mathbf{p}^d(\mathbf{x})=\mathbf{p}(\mathbf{x})-\langle\mathbf{p}\rangle$, can be expressed by integral operators:
\begin{equation}\label{IntegralOp2}
\begin{array}{c}
\displaystyle \boldsymbol{\sigma}^d\mathbf{(x)}=-\left[\boldsymbol{\Lambda}^{(\boldsymbol{\sigma}\dot{\mathbf{U}})}(\mathbf{x};\dot{\mathbf{U}})+\boldsymbol{\Lambda}^{(\boldsymbol{\sigma}\mathbf{E})}(\mathbf{x};\mathbf{E})\right],\\

\displaystyle \mathbf{p}^d\mathbf{(x)}=-\left[\boldsymbol{\Lambda}^{(\mathbf{p}\dot{\mathbf{U}})}(\mathbf{x};\dot{\mathbf{U}})+\boldsymbol{\Lambda}^{(\mathbf{p}\mathbf{E})}(\mathbf{x};\mathbf{E})\right],\\

\end{array} 
\end{equation}
where these integral operators are given in \ref{A}. As in (\ref{Gamma_Operators},   \ref{Consistency3}, \ref{Consistency4}), the consistency conditions may now be expressed as,
\begin{equation}\label{Consistency5}
\mathbf{E}:\left[\mathbf{D}-\mathbf{D}^0\right]^{-1}+\dot{\mathbf{U}}\cdot\boldsymbol{\Lambda}^{(\boldsymbol{\sigma}\dot{\mathbf{U}})}+\mathbf{E}:\boldsymbol{\Lambda}^{(\boldsymbol{\sigma}\mathbf{E})}-\langle\boldsymbol{\sigma}\rangle=\mathbf{0},
\end{equation}
\begin{equation}\label{Consistency6}
\dot{\mathbf{U}}\left[{\nu}-{\nu}^0\right]^{-1}+\dot{\mathbf{U}}\cdot\boldsymbol{\Lambda}^{(\mathbf{p}\dot{\mathbf{U}})}+\mathbf{E}:\boldsymbol{\Lambda}^{(\mathbf{p}\mathbf{E})}-\langle\mathbf{p}\rangle=\mathbf{0}.
\end{equation}
Similarly to (\ref{Functional_1}), it can be shown that the functional,
\begin{equation}\label{Functional_4}
\begin{array}{c}
\displaystyle \mathcal{G}(\hat{\mathbf{E}},\hat{\dot{\mathbf{U}}})\equiv\\

\displaystyle \frac{1}{2}\left[\langle\hat{\mathbf{E}}:\
\{(\mathbf{D}-\mathbf{D}^0)^{-1}+\
\boldsymbol{\Lambda}^{(\boldsymbol{\sigma}{\mathbf{E}})}\}:\hat{\mathbf{E}}\rangle+\langle\hat{\dot{\mathbf{U}}}\cdot\{(\nu-\nu^0)^{-1}+\
\boldsymbol{\Lambda}^{(\mathbf{p}{\dot{\mathbf{U}}})}\}\cdot\hat{\dot{\mathbf{U}}}\rangle\right.\\

\displaystyle \left.+\{\langle\hat{\dot{\mathbf{U}}}\cdot\boldsymbol{\Lambda}^{(\boldsymbol{\sigma}{\dot{\mathbf{U}})}}:\
 \hat{\mathbf{E}}\rangle+\langle\hat{\mathbf{E}}:\
\boldsymbol{\Lambda}^{(\mathbf{p}{\mathbf{E}})}\cdot\hat{\dot{\mathbf{U}}}\rangle\}\right].

\end{array} 
\end{equation}
is real-valued and positive (negative) for any nonzero eigenfields, $\hat{\mathbf{E}}(\mathbf{x})$ and $\hat{\dot{\mathbf{U}}}(\mathbf{x})$, 
when $(\mathbf{D}-\mathbf{D}^0)$ is positive-definite (negative-definite) and 
$(\nu-\nu^0)$ is positive (negative). 
For the \emph{volume average} of the exact eigenfields $\mathbf{E}(\mathbf{x})$ and $\dot{\mathbf{U}}(\mathbf{x})$ 
that satisfy the consistency conditions (\ref{Consistency5}, \ref{Consistency6}) moreover, functional $\mathcal{G}$ becomes
\begin{equation}\label{Functional_5}
\begin{array}{c}
\displaystyle \mathcal{G}(\mathbf{E},\dot{\mathbf{U}})=\
\displaystyle \frac{1}{2}\left[\
\langle\boldsymbol{\sigma}\rangle^*:\
(\bar{\mathbf{D}}-\mathbf{D}^0):\
\langle\boldsymbol{\sigma}\rangle+\
\langle\mathbf{p}\rangle^*\cdot(\bar{\boldsymbol{\nu}}-\nu^0\mathbf{I})\
\cdot{\langle\mathbf{p}\rangle}\right.\\

\left.+\{\langle\boldsymbol{\sigma}\rangle^*:\
\bar{\mathbf{R}}\cdot\
\langle\mathbf{p}\rangle+\
\langle\mathbf{p}\rangle^*\cdot\
\bar{\mathbf{R}}^\dag:\
\langle\boldsymbol{\sigma}\rangle\}\right].
\end{array} 
\end{equation}
Note that the above expression holds only for the \emph{averaged values of the exact homogenizing eigenfields}, $\langle\mathbf{E}\rangle$ 
and $\langle\dot{\mathbf{U}}\rangle$. 
As can be seen,  $\mathcal{G}(\mathbf{E},\dot{\mathbf{U}})$ is the complementary elastodynamic energy of $\Omega^*$ when its effective dynamic elasticity is  $(\bar{\mathbf{D}}-\mathbf{D}^0)$, its effective dynamic density is $(\bar{\boldsymbol{\nu}}-\nu^0\mathbf{I})$, and it is supporting the average stress $\langle\boldsymbol{\sigma}\rangle$ and average momentum $\langle\mathbf{p}\rangle$ at a fixed frequency $\omega$. Similarly to functional (\ref{S_H_Functional_1}), we may consider a functional of eigenstrain and eigenvelocity fields, $\hat{\mathbf{E}}(\mathbf{x}),\hat{\dot{\mathbf{U}}}(\mathbf{x})$, as follows:
\begin{equation}\label{S_H_Functional_2}
\begin{array}{c}
\displaystyle \mathcal{I}(\hat{\mathbf{E}},\hat{\dot{\mathbf{U}}};\langle\boldsymbol{\sigma}\rangle,\langle{\mathbf{p}}\rangle)=\\

\displaystyle \mathcal{G}(\hat{\mathbf{E}},\hat{\dot{\mathbf{U}}})-\
\displaystyle \frac{1}{2}\left[\langle\boldsymbol{\sigma}\rangle:\
\langle\hat{\mathbf{E}}\rangle^*+\
\langle \hat{\mathbf{E}}\rangle:\
\langle\boldsymbol{\sigma}\rangle^*+\
\langle{\mathbf{p}}\rangle\cdot\hat{\dot{\mathbf{U}}}\rangle^*+\
\langle \hat{\dot{\mathbf{U}}}\cdot\langle{\mathbf{p}}\rangle^*\right].
\end{array} 
\end{equation}
Following the arguments concerning the variation of the $\mathcal{J}$ functional, it can be shown that the variation of the $\mathcal{I}$ functional with respect to $\hat{\mathbf{E}}$ and $\hat{\dot{\mathbf{U}}}$ is given by
\begin{equation}
\begin{array}{l}
\displaystyle
[\mathcal{I}(\hat{\mathbf{E}},\hat{\dot{\mathbf{U}}};\langle\boldsymbol{\sigma}\rangle,\langle\mathbf{p}\rangle)]_{\delta\hat{\mathbf{E}}}=\langle\left[\hat{\mathbf{E}}:(\mathbf{D}-\mathbf{D}^0)^{-1}+\hat{\dot{\mathbf{U}}}\cdot\boldsymbol{\Lambda}^{(\boldsymbol{\sigma}{\dot{\mathbf{U}}})}+\hat{\mathbf{E}}:\boldsymbol{\Lambda}^{(\boldsymbol{\sigma}{\mathbf{E}})}-\langle\boldsymbol{\sigma}\rangle\right]:\delta\hat{\mathbf{E}}\rangle;\\\\

\displaystyle
[\mathcal{I}(\hat{\mathbf{E}},\hat{\dot{\mathbf{U}}};\langle\boldsymbol{\sigma}\rangle,\langle\mathbf{p}\rangle)]_{\delta\hat{\dot{\mathbf{U}}}}=\langle\left[\hat{\dot{\mathbf{U}}}({\nu}-{\nu}^0)^{-1}+\hat{\dot{\mathbf{U}}}\cdot\boldsymbol{\Lambda}^{(\mathbf{p}\dot{{\mathbf{U}}})}+\hat{\mathbf{E}}:\boldsymbol{\Lambda}^{(\mathbf{p}{\mathbf{E}})}-\langle\mathbf{p}\rangle\right]:\delta\hat{\dot{\mathbf{U}}}\rangle.
\end{array}
\end{equation}
It can be seen from Eqs. (\ref{Consistency5}, \ref{Consistency6}) that the above variations go to $0$ for the exact eigenstrain and eigenvelocity fields which produce in the equivalent homogeneous solid, the same stress and momentum fields as in the original heterogeneous $\Omega$. Hence, $\mathcal{I}(\mathbf{{E}},\dot{\mathbf{U}};\langle\boldsymbol{\sigma}\rangle,\langle\mathbf{p}\rangle)=-\mathcal{G}(\mathbf{{E}},\dot{\mathbf{U}})$ is the stationary value of Eq. (\ref{S_H_Functional_2}). Moreover, the vanishing of the  variation of $\mathcal{I}(\hat{\mathbf{E}},\hat{\dot{\mathbf{U}}};\langle\boldsymbol{\sigma}\rangle,\langle\mathbf{p}\rangle)$ for arbitrary variations of $\hat{\mathbf{E}}$ and $\hat{\dot{\mathbf{U}}}$ yields the consistency conditions (\ref{Consistency5}, \ref{Consistency6}). Also, it can be shown that for arbitrary eigenfields $\hat{\mathbf{E}}(\mathbf{x}),\hat\dot{\mathbf{U}}(\mathbf{x})$:

\begin{itemize}
\item If $(\mathbf{D}-\mathbf{D}^0)$ is negative-definite (positive-definite) and $(\nu-\nu^0)$
is negative (positive) then
$\mathcal{I}(\mathbf{E},\dot{\mathbf{U}};
\langle\boldsymbol{\sigma}\rangle,\langle\mathbf{p}\rangle)=
-\mathcal{G}(\mathbf{{E}},\dot{\mathbf{U}})$ 
is the maximum (minimum) value of 
$\mathcal{I}(\hat{\mathbf{E}},\hat{\dot{\mathbf{U}}};\langle\boldsymbol{\sigma}\rangle,\langle\mathbf{p}\rangle)$.
\end{itemize}

\subsection{Bounds and inequalities for the $\mathcal{I}$ Functional}

For arbitrary eigenfields which result in field variables which satisfy the generalized elastodynamic Hill condition on the boundary of region $\Omega$, it can be shown that if $\mathbf{D}-\mathbf{D}^0$ is chosen to be negative semidefinite and $\nu-\nu^0$ is chosen to be negative then the following inequality holds:
\begin{equation}\label{Inequality2J}
\begin{array}{c}
\displaystyle \frac{1}{2}\left[\{\langle\boldsymbol{\sigma}:\mathbf{D}:\boldsymbol{\sigma}\rangle+\langle{\mathbf{p}}\cdot{\nu}{\mathbf{p}}\rangle\}-\{\langle\langle\boldsymbol{\sigma}\rangle:\mathbf{D}^0:\langle\boldsymbol{\sigma}\rangle\rangle +\langle\langle{\mathbf{p}}\rangle\cdot{\nu}^0\langle{\mathbf{p}}\rangle\rangle\}\right]\\
\displaystyle +\mathcal{I}(\hat{\mathbf{E}},\hat{\dot{\mathbf{U}}};\langle\boldsymbol{\sigma}\rangle,\langle{\mathbf{p}}\rangle)\leq 0.\\
\end{array} 
\end{equation}
In addition to the above inequality we can now use theorem $D_2$ to derive another inequality analogous to equation (\ref{Inequality_g}). We compare the case of any consistent general boundary data (denoted by the subscript \textbf{g}) and the corresponding uniform velocity boundary data (denoted by the subscript $\mathbf{v}$), for a common average momentum, $\mathbf{p}^0$.   Hence, according to theorem $D_2$ it now follows that,
\begin{equation}\label{Inequality_g2}
\Pi^c_{\mathbf{v}}(\{\mathbf{D}^0-\bar{\mathbf{D}}\},\{\nu^0\mathbf{I}-\bar{\boldsymbol{\nu}}\})\leq\\ 
\mathcal{G}_{\mathbf{g}}(\hat{\mathbf{E}},\hat{\dot{\mathbf{U}}};\langle\boldsymbol{\sigma}\rangle,\langle{\mathbf{p
}}\rangle).
\end{equation}
Here again the right-hand side may be calculated using any suitable operators, for example those corresponding to periodic boundary data.

\section{Conclusions}

In this paper we have shown that the total elastodynamic energy, and the total complementary elastodynamic 
energy of the equivalent solid, when regarded as functionals of the eigenstress (eigenstrain), and eigenmomentum (eigenvelocity), are stationary for the exact eigenstress (eigenstrain), and eigenmomentum (eigenvelocity).This is the dynamic equivalent of the Hashin-Shtrikman variational principle which applies to elastostatic heterogeneous composites. 
In addition, we have developed strict \emph{computable} bounds for these energies that apply to \emph{any spatially variable} (consistent) boundary data. 
In particular Eqs. (\ref{Inequality_g}, \ref{Inequality_g2}) show that the total elastodynamic strain energy and the total elastodynamic complementary energy of the composite can be bounded by considering any general set of integral operators which satisfies (\ref{BD_Zero}). For example, one may use the integral operators given in .

CCC

\section{Acknowledgement}

This research has been conducted at the Center of Excellence for Advanced Materials (CEAM) at the University of California, San Diego, under DARPA AFOSR Grants FA9550-09-1-0709 and RDECOM W91CRB-10-1-0006 to the University of California, San Diego.


\section{References}

\appendix

 \section{Explicit Relations for the Tensors Appearing in the Integral Operators}\label{A}

If we use an isotropic reference material, then we have,

\begin{equation}
C^0_{ijkl}=\lambda^0\delta_{ij}\delta_{kl}+\mu^0[\delta_{ik}\delta_{jl}+\delta_{il}\delta_{jk}]
\end{equation}

\begin{equation}
D^0_{mnij}=\frac{-\lambda^0}{2\mu^0(3\lambda^0+2\mu^0)}\delta_{mn}\delta_{ij}+\frac{1}{4\mu^0}(\delta_{mi}\delta_{nj}+\delta_{mj}\delta_{ni})
\end{equation}
where $\lambda^0,\mu^0$ are Lam$\text{e}^{'}$ constants. 
Moreover, for a periodic unit cell or a finite composite in  parallelepiped  shape, the integral operators in Eq. (\ref{IntegralOp1}) are given by,
\begin{equation}
\begin{array}{c}
\displaystyle \boldsymbol{\Gamma}^{(\boldsymbol{\varepsilon}\mathbf{P})}(\mathbf{x};\mathbf{P})=-\sum_{\boldsymbol{\xi}\neq 0}\frac{1}{\Omega}\int_{\Omega}\boldsymbol{\Gamma}^{(\boldsymbol{\varepsilon}\mathbf{P})}\cdot\mathbf{P}e^{i\boldsymbol{\xi}\cdot(\mathbf{x}-\mathbf{y})}dV_{\mathbf{y}}\\

\displaystyle 
\boldsymbol{\Gamma}^{(\boldsymbol{\varepsilon}\mathbf{\Sigma})}(\mathbf{x};\mathbf{\Sigma})=-\sum_{\boldsymbol{\xi}\neq 0}\frac{1}{\Omega}\int_{\Omega}\boldsymbol{\Gamma}^{(\boldsymbol{\varepsilon}\mathbf{\Sigma})}:\mathbf{\Sigma}e^{i\boldsymbol{\xi}\cdot(\mathbf{x}-\mathbf{y})}dV_{\mathbf{y}}\\

\displaystyle 
\boldsymbol{\Gamma}^{(\dot{\mathbf{u}}\mathbf{P})}(\mathbf{x};\mathbf{P})=-\sum_{\boldsymbol{\xi}\neq 0}\frac{1}{\Omega}\int_{\Omega}\boldsymbol{\Gamma}^{(\dot{\mathbf{u}}\mathbf{P})}\cdot\mathbf{P}e^{i\boldsymbol{\xi}\cdot(\mathbf{x}-\mathbf{y})}dV_{\mathbf{y}}\\

\displaystyle 
\boldsymbol{\Gamma}^{(\dot{\mathbf{u}}\mathbf{\Sigma})}(\mathbf{x};\mathbf{\Sigma})=-\sum_{\boldsymbol{\xi}\neq 0}\frac{1}{\Omega}\int_{\Omega}\boldsymbol{\Gamma}^{(\dot{\mathbf{u}}\mathbf{\Sigma})}:\mathbf{\Sigma}e^{i\boldsymbol{\xi}\cdot(\mathbf{x}-\mathbf{y})}dV_{\mathbf{y}}\\

\end{array} 
\end{equation}
The tensors appearing in the above integrals are given by,
\begin{equation}
\begin{array}{c}
\displaystyle \boldsymbol{\Gamma}^{(\boldsymbol{\varepsilon}\mathbf{P})}=\mathbf{D}^0:\boldsymbol{\Psi};\quad\boldsymbol{\Gamma}^{(\boldsymbol{\varepsilon}\mathbf{\Sigma})}=\boldsymbol{\Pi}-\mathbf{1^{4s}}\\

\displaystyle 
\boldsymbol{\Gamma}^{(\dot{\mathbf{u}}\mathbf{P})}=\mathbf{\Phi};\quad \boldsymbol{\Gamma}^{(\dot{\mathbf{u}}\mathbf{\Sigma})}=\mathbf{\Theta}:\mathbf{D}^0\\

\end{array} 
\end{equation}
where
\begin{eqnarray}\label{APsi}
\Psi_{ijp} & = & -\frac{\omega}{2}\left[\left\{\frac{2c_2^2(c_1^2-c_2^2)}{[\omega^2-c_2^2{\xi}^2][\omega^2-c_1^2{\xi}^2]}\right\}\xi_i\xi_j\xi_p\right. \\
   & + & \left.\left\{\frac{c_1^2-2c_2^2}{\omega^2-c_1^2{\xi}^2}\right\}\delta_{ij}\xi_p+\left\{\frac{c_2^2}{\omega^2-c_2^2{\xi}^2}\right\}\{\xi_i\delta_{jp}+\xi_j\delta_{ip}\}\right]\nonumber
\end{eqnarray}
\begin{eqnarray}\label{AGamma}
\Pi_{mnkl} & = & \frac{1}{\rho^0}\left[\frac{1}{4(\omega^2-c_2^2{\xi}^2)}\{\xi_m\delta_{nk}\xi_l+\xi_m\delta_{nl}\xi_k+\xi_n\delta_{mk}\xi_l+\xi_n\delta_{ml}\xi_k\}\right.
\nonumber \\
   & + & 
\left. \frac{-(c_1^2-2c_2^2)}{2c_2^2(3c_1^2-4c_2^2)}\delta_{mn}\delta_{kl}+\frac{c_1^2-c_2^2}{[\omega^2-c_2^2{\xi}^2][\omega^2-c_1^2{\xi}^2]}\xi_m\xi_n\xi_k\xi_l\right.
 \\
   & + & 
\left.\frac{1}{4c_2^2}\{\delta_{mk}\delta_{nl}+\delta_{ml}\delta_{nk}\}\right]\nonumber
\end{eqnarray}
\begin{equation}\label{APhi}
\Phi_{pj}=\frac{\omega^2}{\rho^0}\left[\frac{c_1^2-c_2^2}{\left[\omega^2-c_1^2\xi^2\right]\left[\omega^2-c_2^2\xi^2\right]}\xi_p\xi_j+\frac{1}{\omega^2-c_2^2\xi^2}\delta_{pj}\right]
\end{equation}
\begin{eqnarray}\label{Theta}
\Theta_{pij} & = & -\frac{\omega}{2}\left[\left\{\frac{2c_2^2(c_1^2-c_2^2)}{[\omega^2-c_2^2{\xi}^2][\omega^2-c_1^2{\xi}^2]}\right\}\xi_i\xi_j\xi_p\right. \\
   & + & \left.\left\{\frac{c_1^2-2c_2^2}{\omega^2-c_1^2{\xi}^2}\right\}\delta_{ij}\xi_p+\left\{\frac{c_2^2}{\omega^2-c_2^2{\xi}^2}\right\}\{\xi_i\delta_{jp}+\xi_j\delta_{ip}\}\right]\nonumber
\end{eqnarray}
In the above expressions $c_1=\sqrt{(\lambda^0+2\mu^0)/\rho^0}$ is the longitudinal wave velocity and $c_1=\sqrt{\mu^0/\rho^0}$ is the shear wave velocity.

Similarly the tensors in Eq. (\ref{IntegralOp2}) are given by
\begin{equation}
\begin{array}{c}
\displaystyle \boldsymbol{\Lambda}^{(\boldsymbol{\sigma}\dot{\mathbf{U}})}=\rho^0\boldsymbol{\Psi};\quad\boldsymbol{\Lambda}^{(\boldsymbol{\sigma}\mathbf{E})}=\mathbf{C}^0:\boldsymbol{\Pi}:\mathbf{C}^0\\

\displaystyle 
\boldsymbol{\Lambda}^{({\mathbf{p}}\dot{\mathbf{U}})}=(\rho^0)^2\mathbf{\Phi}-\mathbf{1^{2}};\quad \boldsymbol{\Lambda}^{({\mathbf{p}}\mathbf{E})}=\rho^0\mathbf{\Theta}\\

\end{array} 
\end{equation}
It may be seen that the tensors appearing in the above equations possess certain symmetries. Specifically we have the following:
\begin{equation}
\begin{array}{c}
\displaystyle {\Gamma}^{(\boldsymbol{\varepsilon}\mathbf{\Sigma})}_{ijkl}={\Gamma}^{(\boldsymbol{\varepsilon}\mathbf{\Sigma})}_{klij}={\Gamma}^{(\boldsymbol{\varepsilon}\mathbf{\Sigma})}_{jikl}={\Gamma}^{(\boldsymbol{\varepsilon}\mathbf{\Sigma})}_{ijlk}\\

\\
\displaystyle {\Gamma}^{(\boldsymbol{\varepsilon}\mathbf{P})}_{ijk}={\Gamma}^{(\dot{\mathbf{u}}\mathbf{\Sigma})}_{kij};\quad{\Gamma}^{(\dot{\mathbf{u}}\mathbf{P})}_{ij}={\Gamma}^{(\dot{\mathbf{u}}\mathbf{P})}_{ji}\\

\end{array} 
\end{equation}
and similar symmetries hold for the $\boldsymbol{\Lambda}$ tensors. These symmetries are essential for the variations of the functionals to have the forms presented in the main text. These symmetries also hold for integral operators corresponding to arbitrary boundary data cases as shown in \cite{willis1980polarizationI,willis1980polarizationII,willis1997dynamics}.

\section{Proof for Identities}\label{B}

For arbitrary eigenfields $\bar{\mathbf{\Sigma}},\bar{\mathbf{P}}$ the disturbance fields $\boldsymbol{\sigma}^d=\boldsymbol{\sigma}-\langle\boldsymbol{\sigma}\rangle$ and $\mathbf{p}^d=\mathbf{p}-\langle\mathbf{p}\rangle$ satisfy 
\begin{equation}
\mathbf{D}^0:(\boldsymbol{\sigma}^d-\bar{\mathbf{\Sigma}})=\boldsymbol{\varepsilon}^d;
\quad \nu^0(\mathbf{p}^d-\bar{\mathbf{P}})=\dot{\mathbf{u}}^d
\label{EeigenstuffInv}
\end{equation}
Moreover, since the eigenfields produce field variables which satisfy (\ref{BD_Zero}) we have
\begin{equation}
\langle\boldsymbol{\sigma}^d:\boldsymbol{\varepsilon}^d\rangle+\langle\mathbf{p}^d\cdot\dot{\mathbf{u}}^d\rangle=0.
\label{EHillUse1}
\end{equation}
It follows from the above equations that the scalar
\begin{equation}
\langle\bar{\mathbf{P}}\cdot\boldsymbol{\Gamma}^{\boldsymbol{\varepsilon}{\mathbf{P}}}:\bar{\mathbf{\Sigma}}+\bar{\mathbf{\Sigma}}:\boldsymbol{\Gamma}^{\boldsymbol{\varepsilon}{\mathbf{\Sigma}}}:\bar{\mathbf{\Sigma}}\rangle+\langle\bar{\mathbf{P}}\cdot\boldsymbol{\Gamma}^{\dot{\mathbf{u}}{\mathbf{P}}}\cdot\bar{\mathbf{P}}+\bar{\mathbf{\Sigma}}:\boldsymbol{\Gamma}^{\dot{\mathbf{u}}{\mathbf{\Sigma}}}\cdot\bar{\mathbf{P}}\rangle
\label{MaxFunct1}
\end{equation}
can be written as
\begin{equation}
\langle\bar{\mathbf{\Sigma}}:\mathbf{D}^0:\bar{\mathbf{\Sigma}}+\bar{\mathbf{P}}\cdot\nu^0\bar{\mathbf{P}}\rangle-\langle\boldsymbol{\sigma}^d:\mathbf{D}^0:\boldsymbol{\sigma}^d+\mathbf{p}^d\cdot\nu^0\mathbf{p}^d\rangle.
\label{MaxFunct2}
\end{equation}
Additionally we have the following; \cite{nemat1999micromechanics}:
\begin{equation}
\begin{array}{c}
\displaystyle \langle\bar{\mathbf{\Sigma}}:(\mathbf{C}-\mathbf{C}^0)^{-1}:\bar{\mathbf{\Sigma}}\rangle=-\langle(\mathbf{D}^0:\bar{\mathbf{\Sigma}}):(\mathbf{D}-\mathbf{D}^0)^{-1}:(\mathbf{D}^0:\bar{\mathbf{\Sigma}})\rangle-\langle\bar{\mathbf{\Sigma}}:\mathbf{D}^0:\bar{\mathbf{\Sigma}}\rangle\\

\displaystyle \langle\bar{\mathbf{P}}\cdot(\rho-\rho^0)^{-1}\bar{\mathbf{P}}\rangle=-\langle(\nu^0\bar{\mathbf{P}})\cdot(\nu-\nu^0)^{-1}(\nu^0\bar{\mathbf{P}})\rangle-\langle\bar{\mathbf{P}}\cdot\nu^0\bar{\mathbf{P}}\rangle\\

\end{array} 
\label{MaxFunct3}
\end{equation}
From Eqs. (\ref{MaxFunct1}, \ref{MaxFunct2}, \ref{MaxFunct3}) it follows that the scalar given by $\mathcal{F}(\bar{\mathbf{\Sigma}},\bar{\mathbf{P}})$ is positive (negative) if $(\mathbf{D}-\mathbf{D}^0)^{-1}$ is negative-definite (positive-definite) and $(\nu-\nu^0)^{-1}$ is negative (positive). Furthermore, the negative-definiteness (positive-definiteness) of $(\mathbf{D}-\mathbf{D}^0)^{-1}$ implies the negative-definiteness (positive-definiteness) of $(\mathbf{D}-\mathbf{D}^0)$ and the positive-definiteness (negative-definiteness) of $(\mathbf{C}-\mathbf{C}^0)$. Therefore, the functional in Eq. (\ref{IFunctionalMod}) assumes a maximum or a minimum value depending upon the choice of the reference material.

When a reference material is chosen such that $\mathbf{C-C}^0$ is negative-semidefinite and ${\rho}-{\rho}^0$ is negative then for any arbitrary strain fields, $\boldsymbol{\varepsilon}$ and $\hat{\boldsymbol{\varepsilon}}$, and any arbitrary velocity fields, $\dot{\mathbf{u}}$ and $\hat{\dot{\mathbf{u}}}$, the following inequality holds:
\begin{equation}\label{Inequality1B}
\frac{1}{2}\left[\langle(\boldsymbol{\varepsilon}-\hat{\boldsymbol{\varepsilon}}):(\mathbf{C-C}^0):(\boldsymbol{\varepsilon}-\hat{\boldsymbol{\varepsilon}})\rangle+\langle(\dot{\mathbf{u}}-\hat{\dot{\mathbf{u}}})({\rho}-{\rho}^0)\cdot(\dot{\mathbf{u}}-\hat{\dot{\mathbf{u}}})\rangle\right]\leq 0.
\end{equation}
For arbitrary eigenstress and eigenmomentum fields, $\hat{\mathbf{\Sigma}}$ and $\hat{{\mathbf{P}}}$, and given average strain and velocity fields, $\langle\boldsymbol{\varepsilon}\rangle$ and $\langle\dot{\mathbf{u}}\rangle$, we consider the following strain and velocity fields:
\begin{equation}\label{defineStressMomentumB}
\begin{array}{l}

\displaystyle \boldsymbol{\varepsilon}=\langle\boldsymbol{\varepsilon}\rangle-\mathbf{\Gamma}^{(\boldsymbol{\varepsilon}\hat{\mathbf{\Sigma}}\hat{{\mathbf{P}}})};\quad \hat{\boldsymbol{\varepsilon}}=(\mathbf{C-C}^0)^{-1}:\hat{\mathbf{\Sigma}},\\

\displaystyle \dot{\mathbf{u}}=\langle\dot{\mathbf{u}}\rangle-\mathbf{\Gamma}^{(\dot{\mathbf{u}}\hat{\mathbf{\Sigma}}\hat{{\mathbf{P}}})};\quad \hat{\dot{\mathbf{u}}}=({\rho}-{\rho}^0)^{-1}:\hat{\mathbf{P}},\\

\end{array} 
\end{equation}
where $\mathbf{\Gamma}^{(\boldsymbol{\varepsilon}\hat{\mathbf{\Sigma}}\hat{{\mathbf{P}}})}=\hat{{\mathbf{P}}}\cdot\boldsymbol{\Gamma}^{(\boldsymbol{\varepsilon}\mathbf{P})}+\hat{\mathbf{\Sigma}}:\boldsymbol{\Gamma}^{(\boldsymbol{\varepsilon}{\mathbf{\Sigma}})}$ and $\mathbf{\Gamma}^{(\dot{\mathbf{u}}\hat{\mathbf{\Sigma}}\hat{{\mathbf{P}}})}=\hat{\mathbf{P}}\cdot\boldsymbol{\Gamma}^{(\dot{\mathbf{u}}\mathbf{P})}+\hat{\mathbf{E}}:\boldsymbol{\Gamma}^{(\dot{\mathbf{u}}{\mathbf{E}})}$. Since $-\mathbf{\Gamma}^{(\boldsymbol{\varepsilon}\hat{\mathbf{\Sigma}}\hat{{\mathbf{P}}})}$ and $-\mathbf{\Gamma}^{(\dot{\mathbf{u}}\hat{\mathbf{\Sigma}}\hat{{\mathbf{P}}})}$ are zero volume average deviatoric parts of the strain and the velocity fields, the strain and velocity fields, $\boldsymbol{\varepsilon}=\langle\boldsymbol{\varepsilon}\rangle-\mathbf{\Gamma}^{(\boldsymbol{\varepsilon}\hat{\mathbf{\Sigma}}\hat{{\mathbf{P}}})}$ and $\dot{\mathbf{u}}=\langle\dot{\mathbf{u}}\rangle-\mathbf{\Gamma}^{(\dot{\mathbf{u}}\hat{\mathbf{\Sigma}}\hat{{\mathbf{P}}})}$, therefore, are such that,
\begin{equation}
\langle\boldsymbol{\varepsilon}\rangle=\langle\langle\boldsymbol{\varepsilon}\rangle-\mathbf{\Gamma}^{(\boldsymbol{\varepsilon}\hat{\mathbf{\Sigma}}\hat{{\mathbf{P}}})}\rangle=\langle\boldsymbol{\varepsilon}\rangle ;\quad
\langle\dot{\mathbf{u}}\rangle=\langle\langle\dot{\mathbf{u}}\rangle-\mathbf{\Gamma}^{(\dot{\mathbf{u}}\hat{\mathbf{\Sigma}}\hat{{\mathbf{P}}})}\rangle=\langle\dot{\mathbf{u}}\rangle.
\end{equation}
Since the fields satisfy the generalized elastodynamic hill condition we also have the following:
\begin{equation}\label{zeroVolumeAverage1}
\begin{array}{c}

\displaystyle
\langle\boldsymbol{\varepsilon}^d:\boldsymbol{\sigma}^d\rangle+\langle\dot{\mathbf{u}}^d\cdot\mathbf{p}^d\rangle=\\

\displaystyle \langle\mathbf{\Gamma}^{(\boldsymbol{\varepsilon}\hat{\mathbf{\Sigma}}\hat{\mathbf{P}})}:\left[\mathbf{C}^0:\mathbf{\Gamma}^{(\boldsymbol{\varepsilon}\hat{\mathbf{\Sigma}}\hat{\mathbf{P}})}+\hat{\mathbf{\Sigma}}\right]\rangle+\langle\mathbf{\Gamma}^{(\dot{\mathbf{u}}\hat{\mathbf{\Sigma}}\hat{\mathbf{P}})}\cdot\left[{\rho}^0\mathbf{\Gamma}^{(\dot{\mathbf{u}}\hat{\mathbf{\Sigma}}\hat{\mathbf{P}})}+\hat{\mathbf{P}}\right]\rangle=0.\\

\end{array} 
\end{equation}
Substituting $\hat{\boldsymbol{\varepsilon}}$ and $\hat{\dot{\mathbf{u}}}$ from Eq. (\ref{defineStressMomentumB}) into Eq. (\ref{Inequality1B}) we have
\begin{equation}
\begin{array}{c}

\displaystyle
\frac{1}{2}\left[\langle\boldsymbol{\varepsilon}:\mathbf{(C-C}^0):\boldsymbol{\varepsilon}\rangle+\langle\hat{\mathbf{\Sigma}}:(\mathbf{C-C}^0)^{-1}:\hat{\mathbf{\Sigma}}\rangle-\langle\boldsymbol{\varepsilon}:\hat{\mathbf{\Sigma}}\rangle-\langle\hat{\mathbf{\Sigma}}:\boldsymbol{\varepsilon}\rangle\right.\\

\displaystyle +\left.\langle\dot{\mathbf{u}}\cdot({{\rho}-{\rho}}^0)\dot{\mathbf{u}}\rangle+\langle\hat{\mathbf{P}}\cdot({{\rho}-{\rho}}^0)^{-1}\hat{\mathbf{P}}\rangle-\langle\dot{\mathbf{u}}\cdot\hat{\mathbf{P}}\rangle-\langle\hat{\mathbf{P}}\cdot\dot{\mathbf{u}}\rangle\right]\leq 0\\

\end{array} 
\end{equation}
Now substituting $\boldsymbol{\varepsilon}=\langle\boldsymbol{\varepsilon}\rangle-\mathbf{\Gamma}^{(\boldsymbol{\varepsilon}\hat{\mathbf{\Sigma}}\hat{\mathbf{P}})}$ and $\dot{\mathbf{u}}=\langle\dot{\mathbf{u}}\rangle-\mathbf{\Gamma}^{(\dot{\mathbf{u}}\hat{\mathbf{\Sigma}}\hat{\mathbf{P}})}$ in the above equation and using Eq. (\ref{zeroVolumeAverage1}), it can be shown that the above equation can be written in the following form:
\begin{equation}\label{Inequality2B}
\begin{array}{c}

\displaystyle
\frac{1}{2}\left[\{\langle\boldsymbol{\varepsilon}:\mathbf{C}:\boldsymbol{\varepsilon}\rangle+\langle\dot{\mathbf{u}}\cdot{\rho}\dot{\mathbf{u}}\rangle\}-\{\langle\langle\boldsymbol{\varepsilon}\rangle:\mathbf{C}^0:\langle\boldsymbol{\varepsilon}\rangle\rangle +\langle\langle\dot{\mathbf{u}}\rangle\cdot{\rho}^0\langle\dot{\mathbf{u}}\rangle\rangle\}\right]\\

\displaystyle +\mathcal{J}(\hat{\mathbf{\Sigma}},\hat{\mathbf{P}};\langle\boldsymbol{\varepsilon}\rangle,\langle\dot{\mathbf{u}}\rangle)\leq 0.\\

\end{array} 
\end{equation}
The above is the proof for Eq. (\ref{Inequality2}). Analogous proofs for the $\mathcal{I}$ functional can be derived similarly.

\end{document}